\documentclass[fleqn,twoside]{article}
\usepackage{graphicx}
\usepackage{espcrc2}

\def\text#1{{\rm #1}}

\def\bm{{\bf m}}
\def\bn{{\bf n}}

\def\Tr{\text{Tr}\,}

\title{Towards a strong-coupling theory of QCD at finite density\thanks{
Presented at Lattice 2002, Cambridge, MA, USA, June 2002}}
\author{B. Bringoltz
and B. Svetitsky\address{School of Physics and Astronomy,
Raymond and Beverly Sackler Faculty of Exact Sciences,
Tel Aviv University,
69978 Tel Aviv, Israel}}
\begin{document}
\begin{abstract}
We apply strong-coupling perturbation theory to the QCD lattice
Hamiltonian.
We begin with naive, nearest-neighbor fermions and subsequently break
the doubling symmetry with next-nearest-neighbor terms.
The effective Hamiltonian is that of an antiferromagnet with an added
kinetic term for baryonic ``impurities,'' reminiscent of the $t$--$J$
model of high-$T_c$ superconductivity.
As a first step, we fix the locations of the baryons and make them
static.
Following analyses of the $t$--$J$ model, we apply large-$N$ methods
to obtain a phase diagram in the $N_c,N_f$ plane at zero temperature
and baryon density.
Next we study a simplified $U(3)$ toy model, in which
we add baryons to the vacuum.
We use a coherent state formalism to write a path
integral which we analyze with mean field theory, obtaining a phase diagram in
the $(n_B,T)$ plane.

\vspace{1pc}
\end{abstract}

\maketitle

Color superconductivity \cite{Krishna} at high density is so far a prediction
only of weak-coupling analysis, valid (if at all) only at very
high densities.
We seek confirmation from methods that do not depend on weak
coupling, as well as an extension to the regime of moderate
densities.
Since Euclidean Monte Carlo methods are unavailable when the
chemical potential is non-zero, we turn to the strong-coupling limit
of QCD; harking back to the early days of lattice gauge theory, we
work in the Hamiltonian formulation.
We derive an effective Hamiltonian for color-singlet states that
takes the form of an antiferromagnet with a kinetic term for
baryons.
This effective Hamiltonian is very difficult to study.
As a first step, we fix the position of the baryons and study
mesonic excitations in the baryonic background.
Coherent-state methods then enable us to derive equivalent models
that are tractable in various limits of large $N_c$ and/or $N_f$.
We benefit from the considerable work done on antiferromagnets
in the context of large-$T_c$ superconductors \cite{MA,AA,RS}, as well as from
the early work of the SLAC group \cite{SLAC} and of Smit \cite{Smit} on
strong-coupling QCD.

\section{The effective Hamiltonian}

The lattice gauge Hamiltonian is composed of electric, magnetic, and
fermion terms,
\begin{equation}
H=H_E+H_U+H_F,
\end{equation}
where the first term is the unperturbed Hamiltonian in the strong coupling
limit,
\begin{equation}
H_E=\frac12g^2\sum_{\bn\mu}E_{\bn\mu}^2.
\end{equation}
The ground state sector of $H_E$ is highly degenerate, consisting of
of all states with zero electric flux, whatever their fermion content,
\begin{equation}
|0\rangle|\chi\rangle_F=\left[\prod_{\bn\mu}|E_{\bn\mu}^2=0\rangle\right]|\chi
\rangle_F.
\end{equation}
Neglecting the magnetic term, which only contributes in high order,
we perturb with the fermion kinetic term,
\begin{eqnarray}
H_F&=&-i\sum_{\bn\mu}\psi^{\dag}_\bn\alpha_\mu \nonumber\\
&&\quad\times\sum_{j>0}D(j)\left(\prod U_{\bn+k\hat\mu,\mu}\right)
\psi_{\bn+j\mu}
\end{eqnarray}
We use four-component fermions with a general (diagonal) kernel
$D(j)$.
This is the best we can do, since domain-wall fermions are unavailable to
us at strong coupling \cite{BS} and there is no Hamiltonian overlap formalism
\cite{CHN}.
We will discuss the properties of these fermions in a moment.

The degeneracy is lifted in second order via diagonalization of
an effective Hamiltonian,
\begin{eqnarray}
H_{\text{eff}}^{(2)}&=&
\sum_{\bn\mu j} K(j)
\left(s_\eta^\mu\right)^{j+1}
\left(\psi^{\dag}M^\eta\psi\right)_\bn\nonumber\\
&&\qquad\times\left(\psi^{\dag}M^\eta\psi\right)_{\bn+j\hat\mu},
\label{Heff2}
\end{eqnarray}
where $s_\eta^\mu$ is a sign factor and $K(j)$ is a new kernel.
The effective Hamiltonian contains fermion bilinears at each site,
\begin{equation}
Q^\eta_\bn=\left(\psi^{\dag}M^\eta\psi\right)_\bn,
\end{equation}
where the $M^\eta$ matrices act on the Dirac and the flavor indices,
$M^\eta=\Gamma^A\otimes\lambda^a$.
The operators $Q^\eta_\bn$ generate a $U(4N_f)$ algebra.
If we fix the baryon number $B$ on a site,
the color-singlet states on that site make up an irreducible representation
of this algebra, whose Young tableau has $m=B+2N_f$ rows and $N_c$ columns
(see Fig.~1).

\begin{figure}[htb]
\includegraphics[scale=0.5]{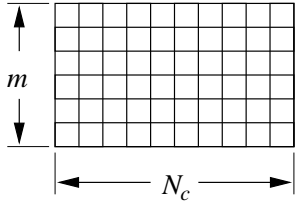}
\caption{Young tableau for single-site states with $B=m-2N_f$.}
\end{figure}

Expressed in terms of $Q^\eta_\bn$, the effective Hamiltonian is an
antiferromagnet.
If the kernel $D(j)$ is chosen to give nearest-neighbor couplings only,
$H_{\text{eff}}^{(2)}$ will take the form
\begin{equation}
H_{\text{eff}}^{(2)} =K(1) \sum_{\bn\mu} Q^\eta_\bn Q^\eta_{\bn+\hat\mu}.
\label{Hnn}
\end{equation}
This antiferromagnet has the accidental $U(4N_f)$ symmetry of
naive fermions, which is responsible for part of the doubling problem.
The even-$j$ terms in (\ref{Heff2}) break this symmetry to
$SU(N_f)\times SU(N_f)\times U(1)_V\times U(1)_A$, which [apart from the
unbreakable axial $U(1)$] is the desired continuum symmetry.
Since we are interested in strong coupling where the free fermion dispersion
relation is of no interest, this might be a good-enough partial solution of
the doubling problem.

The theory of $H_{\text{eff}}^{(2)}$ contains only static baryons, which
make their presence felt through fixing the rep of $U(4N_f)$ on each site.
These baryons can move in the next order in perturbation theory (only if
$N_c=3$---a fortunate special case).
The effective Hamiltonian in third order is
\begin{eqnarray}
H_{\text{eff}}^{(3)}&=&
-i\sum_{j\bn \mu } \tilde K(j) b_\bn^{\dag I}\left[
\alpha_\mu\otimes\alpha_\mu\otimes\alpha_\mu\right]_{II'}b_{\bn+j\hat\mu}^{I'}
\nonumber\\
&&\quad + h.c.,
\end{eqnarray}
where the baryon operators are the color singlets
$b^I=\epsilon_{\alpha\beta\gamma}\psi_\alpha^a\psi_\beta^b\psi_\gamma^c$
that belong to the 
\includegraphics[scale=0.25]{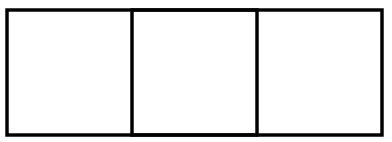}
rep of $U(4N_f)$.
The simple form of $H_{\text{eff}}$ is deceptive, for the $b^I$'s are composite
and hence do not obey canonical anticommutation relations.

For nearest-neighbor fermions, the usual spin diagonalization \cite{SLAC}
gives a simplified Hamiltonian,
\begin{equation}
H_{\text{eff}}^{(3)}=-i\tilde K(1)
\sum_{\bn\mu}b_\bn^{\dag I}b_{\bn+j\hat\mu}^I
\eta_\mu(\bn) + h.c.
\end{equation}
The complete Hamiltonian $H_{\text{eff}}^{(2)}+H_{\text{eff}}^{(3)}$ resembles
that of the $t$--$J$ model, which represents the strong-binding limit of
the Hubbard model and is much studied in connection with high-$T_c$
superconductivity \cite{Auerbach}.
But our Hamiltonian is much more complex.

\section{Static baryons}

Let us beat a strategic retreat to the second-order theory, where baryons
constitute a static background.
If we begin with the nearest-neighbor model, in a state with {\em no\/}
baryons, then the effective Hamiltonian (\ref{Hnn}) is that of a $U(4N_f)$
antiferromagnet with spins in a rep specified by $N_c$ and by $m=2N_f$
(which can vary from site to site).
This can be studied in the limits of large $N_f,N_c$ by various transformations
\cite{MA,AA,RS,Auerbach}, and the result---still for $B=0$---is the
phase diagram in Fig.~2.
\begin{figure}[htb]
\includegraphics[scale=0.35]{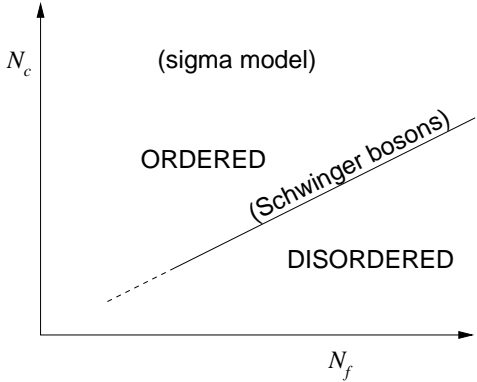}
\caption{Phase diagram for the nn antiferromagnet at $T=0$}
\end{figure}
(Note that
the coupling constant $K(1)$ is just a scale, which has no effect at $T=0$.)
The large-$N_f$ phase is disordered, and will not concern us.
The location of the phase boundary can be established by studying
a Schwinger boson representation of the $U(4N_f)$ spins, and its slope
turns out to be $N_c/N_f=0.31$; this means that the QCD vacuum
is safely in the {\em ordered\/} phase for any reasonable number of flavors.

The ordered phase is conveniently studied in a $\sigma$ model representation,
which comes from rewriting (\ref{Hnn}) in a basis of spin coherent states
\cite{RS}.
(This is valid for any $N_c,N_f$ but proves soluble in the $N_c\to\infty$
limit.)
The degrees of freedom of the $\sigma$ model are the $N\times N$ matrices
(with $N\equiv4N_f$)
\begin{equation}
Q_\bn=U_\bn\Lambda U_\bn^{\dag},
\end{equation}
where
\begin{equation}
\Lambda=\left(
\begin{array}{cc}
1_{N/2}&0\\
0&-1_{N/2}
\end{array}\right).
\end{equation}
The field $U_\bn$ runs over the group $U(N)$, and
the manifold covered by $Q_\bn$ is the coset space $U(N)/[U(N/2)\times U(N/2)]$.
The action of the $\sigma$ model (in continuous time) is
\begin{eqnarray}
S=\int_0^\beta d\tau\left[\frac{N_c}2\sum_{\bn}\Tr\Lambda U^{\dag}_{\bn}
\partial_\tau U_{\bn}
-H(Q(\tau))\right],&&\nonumber\\\ 
\label{action}
\end{eqnarray}
where, in terms of the matrices $Q_\bn$, the nearest-neighbor Hamiltonian
takes the form
\begin{equation}
H=\left(\frac{N_c}2\right)^2 K(1) \sum \Tr Q_\bn Q_{\bn+\hat\mu}.
\end{equation}
Clearly as $N_c\to\infty$ the ground state is the classical minimum
of $H$, in which on all sites $Q_\bn=Q_0$, which can be rotated to
$Q_\bn=\Lambda$.
Thus the symmetry is spontaneously broken as
$U(N)\to U(N/2)\times U(N/2)$, with $N^2/2$ Goldstone bosons \cite{Smit}.

We now restore next-nearest-neighbor couplings, viz.
\begin{equation}
H'=\frac{K(2)}{2} \left( \frac{N_c}{2} \right)^2 \sum_{\bn\mu} Q^\eta_{\bn}Q^\eta_{\bn+2\hat\mu}s^\mu_\eta,
\end{equation}
where $Q^{\eta}_{\bn}=2 \Tr M^\eta Q_\bn$.
The symmetry of the theory, as discussed above, is $U(N_f)\times U(N_f)$;
the classical minimum is at $Q_\bn=\gamma_0$, which breaks all the axial
generators and leaves the vector $U(N_f)$ unbroken.
This is what we would expect for the ground state in the vacuum sector.

\section{Adding baryons}

The $B=0$ states considered above were specified by choosing the $m=2N_f$
rep on each site.
Choosing a different rep adds (or subtracts) baryons on a site-by-site
basis.
For instance, we can add a single baryon by adding a row to the Young
tableau (Fig.~3).
\begin{figure}[htb]
\includegraphics[scale=0.5]{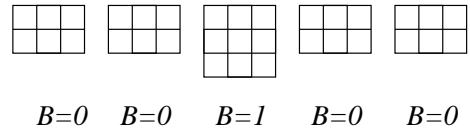}
\caption{Adding one baryon to a site in the $N_f=1$ theory
[a $U(4)$ antiferromagnet]}
\end{figure}
One can similarly add baryons on an entire sublattice.
The $N_c\to\infty$ limit directs us to find the classical ground state of
the Hamiltonian, which always breaks the symmetry spontaneously along the
lines shown in the section above.
We can study the effects of the fluctuations by doing mean field theory
for finite $N_c$.
To do this, we drop the kinetic term in (\ref{action}) and go to $T\not=0$
by calculating the resulting classical partition function.

\section{Mean field theory} 

In MF theory, we write down a trial Hamiltonian and calculate a
variational free energy $\Phi$.
The simplest trial Hamiltonian, containing no site--site correlations, is
\begin{equation}
H_0 = \sum_\bn \vec{Q}_\bn \cdot \vec{h}_\bn,
\end{equation}
where the magnetic fields $\vec{h}_\bn$ are variational parameters.
(We write $\vec{Q}_\bn$ for the vector whose $N^2$ components are
$Q^\eta_\bn$.)\ 
The free energy obeys $F \leq \Phi$ with
\begin{eqnarray}
\Phi &=& -\sum_\bn \log
\left ( \int dQ_\bn \, e^{-K \vec{Q}_\bn \cdot \vec{h}_\bn}
\right ) \nonumber \\
&&\quad+ K\sum_{\langle \bm\bn \rangle}
\vec{\mu}_\bm \cdot \vec{\mu}_\bn - 
\sum_\bn \vec{\mu}_\bn \cdot \vec{h}_\bn
\end{eqnarray}
Here $K=\beta \frac{K(1)}{2} \left( \frac{N_c}{2} \right)^2$ and $\vec{\mu}$
is the magnetization,
\begin{equation}
\vec{\mu}_\bn =
\frac{\int dQ_\bn\,
e^{-\vec{Q}_\bn \cdot \vec{h}_\bn} \vec{Q}_\bn}
{\int dQ_\bn\,e^{-\vec{Q}_\bn \cdot \vec{h}_\bn}}.
\label{eq:magnetization}
\end{equation}
Note that the integration measure $dQ_\bn$ depends on the $U(N)$
representation chosen for site $\bn$.

We minimize $\Phi$ with respect to $\{ \vec{h}_\bn \}$ and get the mean field equations,
\begin{equation}
\vec\mu_\bn = \frac{\int dQ_\bn\,e^{-K\vec{Q}_\bn \cdot
\sum_{\bm(\bn)} \vec{\mu}_\bm} \vec{Q}_\bn}
{\int dQ_\bn\,e^{-K\vec{Q}_{\bn} \cdot \sum_{\bm(\bn)}
\vec{\mu}_\bm}}.
\end{equation}
If all sites are in the same rep of $U(N)$, then the bipartite nature
of the antiferromagnetic system gives two coupled sets of MF equations.
Otherwise one gets as many coupled MF equations as there are inequivalent
sites.

After solving for $\vec{\mu}_\bn$ we evaluate $\Phi$ via
\begin{eqnarray}
\Phi &=& -\sum_\bn \log
\left ( \int dQ \, e^{-K \vec{Q}_\bn \cdot \sum_{\bm(\bn)}
\vec{\mu}_\bm} \right ) \nonumber \\
&&  \quad-K\sum_{\langle \bm\bn \rangle} \vec{\mu}_\bn \cdot \vec{\mu}_\bm.
\end{eqnarray}
We seek the global minimum of $\Phi$.
Once we find it, we can examine its symmetry properties and identify the phase
favored at temperature $T$.

\section{$U(3)$, a toy model}

The $N_f=1$ theory, the $U(4)$ antiferromagnet (with impurities), contains many
degrees of freedom in which to do mean field theory.
A simpler non-trivial model is a toy model with $U(3)$ symmetry,
which does not correspond to an actual value of $N_f$.
The most symmetric $B=0$ state in this model cannot have $B=0$ on every site,
but must alternate between the reps corresponding to $B=\pm1/2$ (see Fig.~4).
\begin{figure}[htb]
\includegraphics[scale=0.5]{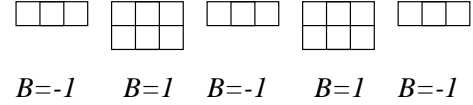}
\caption{The $B=0$ state in the $U(3)$ toy model}
\end{figure}
A $B\not=0$ state is specified by breaking the alternating pattern, as in
Fig.~5.
\begin{figure}[htb]
\includegraphics[scale=0.5]{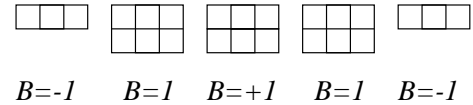}
\caption{Adding a baryon in the $U(3)$ toy model}
\end{figure}
We in fact create a non-zero density of baryons by adding a row to one or more
sublattices, forming a lattice with a unit cell of $2\times2\times2$ sites.
The unit cell may contain sites with $m=1$,~2,~or even~3.
For $m=1,2$ the manifold of $\vec Q_\bn$ is $U(3)/[U(2)\times U(1)]$,
while for $m=3$, the singlet state, we have $\vec Q_\bn=0$.

The form of $U_{\bn}$ for a site with $m=2$ is $U_{\bn}=e^{A_{\bn}}$, with 
\begin{equation}
A = \left( \begin{array}{ccc}
        0       &       0       &       a_1     \\
        0       &       0       &       a_2     \\
        -a^{*}_1 &      -a^{*}_2 &      0       \end{array}     \right).
\end{equation}
We write
\begin{eqnarray}
a_1 &=& \rho \cos (\theta/2) e^{i\alpha_1}   \nonumber       \\
a_2 &=& \rho \sin (\theta/2) e^{i\alpha_2},
\end{eqnarray}
with $0<\rho,\theta<\pi$ and $0<\alpha_{1,2}<2\pi$.
With these definitions, $\vec{Q}$ is given by
\begin{equation}
\vec{Q} = -2\left(
\begin{array}{c}
-\frac1{\sqrt6}     \\
\sin^2 \rho \sin \theta \cos \phi        \\
-\sin^2 \rho \sin \theta \sin \phi      \\
\sin^2 \rho \cos \theta                       \\
\sin 2\rho \cos \theta/2 \cos\left(\frac12(\psi+\phi)\right)     \\
-\sin 2\rho \cos \theta/2 \sin\left(\frac12(\psi+\phi)\right)    \\
\sin 2\rho \sin \theta/2 \cos\left(\frac12(\psi-\phi)\right)    \\
-\sin 2\rho \sin \theta/2 \sin\left(\frac12(\psi-\phi)\right)    \\
-\sqrt3\left(\cos^2 \rho -\frac13\right)
\end{array}     \right)
\label{Q_form}
\end{equation}
Here $\phi \equiv \alpha_1-\alpha_2$ and $\psi = \alpha_1 + \alpha_2$, with $0<\psi<4\pi$ and $0<\phi<2\pi$.
The induced measure on the four-dimensional manifold turns out to be
\begin{equation}
dQ = (1-\cos^2 \rho)\cos\rho\,d(\cos\rho) d(\cos\theta)\frac{d\phi}{2\pi}
\frac{d\psi}{4\pi}
\end{equation}

An example:
For the $B=0$ case shown in Fig.~4, the inequivalent sites are just the
even and odd sublattices.
The corresponding MF equations are
\begin{eqnarray}
\vec\mu_1 &=&  \frac{\int dQ\,e^{-6K\vec{Q} \cdot \vec{\mu}_2} \vec{Q}}
{\int dQ\,e^{-6K\vec{Q} \cdot \vec{\mu}_2}}      \\      \nonumber
\vec\mu_2 &=& -\frac{\int dQ\,e^{+6K\vec{Q} \cdot \vec{\mu}_1} \vec{Q}}
{\int dQ\,e^{+6K\vec{Q} \cdot \vec{\mu}_1}}.
\end{eqnarray}
Cases with $B\not=0$ will give more coupled sets of MF equations, according
to the number of inequivalent sites in the unit cell.

For each baryonic configuration we obtain a phase transition as a function
of temperature.
In some cases it is
first order and in others, second order.
In all cases the symmetry breakdown at low temperature is
$U(3) \to U(3)/[U(2)\times U(1)]$.
As we increase the baryon density the transition temperature decreases,
but it never vanishes; MF theory always breaks symmetry at $T=0$,
even in one dimension.
We summarize our findings in the phase diagram in the temperature--density
plane shown in Fig.~6.

\begin{figure*}[htb]
\includegraphics[scale=0.75]{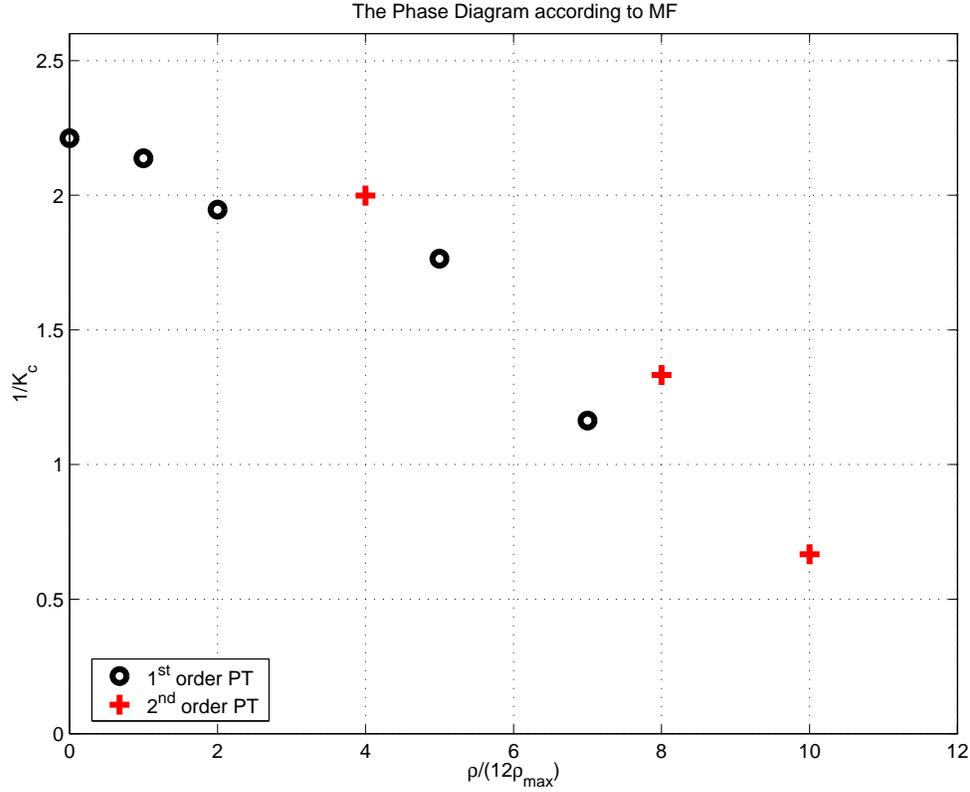}
\caption{Classical phase transition of the $U(3)$ antiferromagnet
with a density of baryon impurities.
The ordinate is proportional to temperature, while the
abscissa is baryon density in units of the maximum saturation density
$\rho_{\rm max}$.}
\end{figure*}

The future holds, we hope, the removal of the various approximations
that led from QCD to the $U(3)$ toy model.
To begin with, we must do better than classical mean field theory and
include the quantum kinetic term in the $\sigma$ model (\ref{action}).
The $U(3)$ model must be generalized to the $U(4N_f)$ symmetry
group of naive fermions, which should be broken to $U(N_f)\times U(N_f)$
by the nnn coupling.
Going beyond the static baryon picture, we can disorder the baryon
background by a replica method; eventually dynamical baryons should
be included with the third-order kinetic term.
Once the theory becomes realistic enough, we can compare the results at
each step to the weak-coupling predictions of color superconductivity
 with various values of $N_f$ \cite{Schafer}.

There is, however, a limitation inherent in the strong-coupling theory.
As we saw in the mean-field analysis above, one is easily misled into
filing the lattice with baryons to saturation.
The baryon density is limited according to
\begin{equation}
\frac BV<\frac{2N_f}{a^3},
\end{equation}
and strong coupling means large lattice spacing $a$.
It is possible that color superconductivity will not show up at any
density short of saturation.
In that case we will have to content ourselves with a new effective theory
for baryonic matter, short of the transition to quark matter.

\end{document}